\documentclass[a4paper,11pt]{article}
\pdfoutput=1 

\usepackage{jheppub} 

\usepackage[T1]{fontenc} 
\usepackage{mathtools}
\usepackage{physics}

\usepackage{xcolor}
\usepackage{ulem}
\usepackage{cancel}

\graphicspath{{images/}}

\title{\boldmath Ultralight Millicharged Dark Matter via Misalignment}

\author[a]{Zachary Bogorad}
\author[b]{and Natalia Toro}

\affiliation[a]{Stanford Institute for Theoretical Physics, Stanford University, Stanford, CA 94305, USA}
\affiliation[b]{SLAC National Accelerator Laboratory, 2575 Sand Hill Road, Menlo Park, CA 94025, USA}

\emailAdd{zbogorad@stanford.edu}
\emailAdd{ntoro@slac.stanford.edu}

\abstract{We explore the cosmology and phenomenology of millicharged and millicharge-like dark matter with masses from 1 eV to 10 keV and charges of $10^{-18}$ to $10^{-14}$.  Dark matter in this mass range cannot be thermally produced, but can arise from non-thermal mechanisms.  We propose a concrete model employing a spontaneously broken approximate global symmetry, in which millicharged dark matter is produced via the misalignment mechanism.  We show that this production mechanism is cosmologically consistent and compatible with the observed dark matter abundance. This model can be implemented using either fundamental scalars or hidden-sector quarks, and coupled either to the Standard Model photon or to a hidden photon. We then consider the phenomenology of light millicharged dark matter, regardless of its cosmological origin, and determine the parameter space consistent with existing experiments and observations. A significant part of the new parameter space we consider may be accessible in the near future through direct deflection experiments, measurements of the cosmic microwave background blackbody spectrum, and future constraints on plasma instabilities due to dark matter self-interaction.}

\begin{document} 
\maketitle
\flushbottom

\section{Introduction}

Though dark matter makes up the majority of the matter in the universe, its nature remains a mystery. Early dark matter searches were largely focused on GeV-scale WIMP dark matter, motivated by the so-called ``WIMP miracle'' \cite{Arcadi2018}. However, as WIMP detectors have now ruled out a large fraction of the best motivated WIMP parameter space, alternative models of dark matter are gaining increasing attention. These models span an extraordinary range of properties, including masses between roughly $10^{-22}$ eV and several solar masses. An enormous variety of experiments are now being operated or designed to cover these possibilities \cite{battaglieri2017cosmic}.

One potential category of light dark matter models is millicharged dark matter \cite{Davidson_2000}. True millicharges are dark matter particles with charges under the Standard Model $U(1)_{\rm EM}$ gauge group. These charges are typically much smaller than the electric charge in order to avoid existing constraints on charged dark matter. Alternatively, dark matter may couple to a very light hidden photon that is then kinetically mixed with the Standard Model photon. In this case, the hidden photon mediates millicharge-like interactions between dark matter and EM charges. Such “effective millicharges” are phenomenologically similar to true millicharges in many—but not all—contexts.

\begin{figure}[t]
\centering
\includegraphics[width=0.6\linewidth]{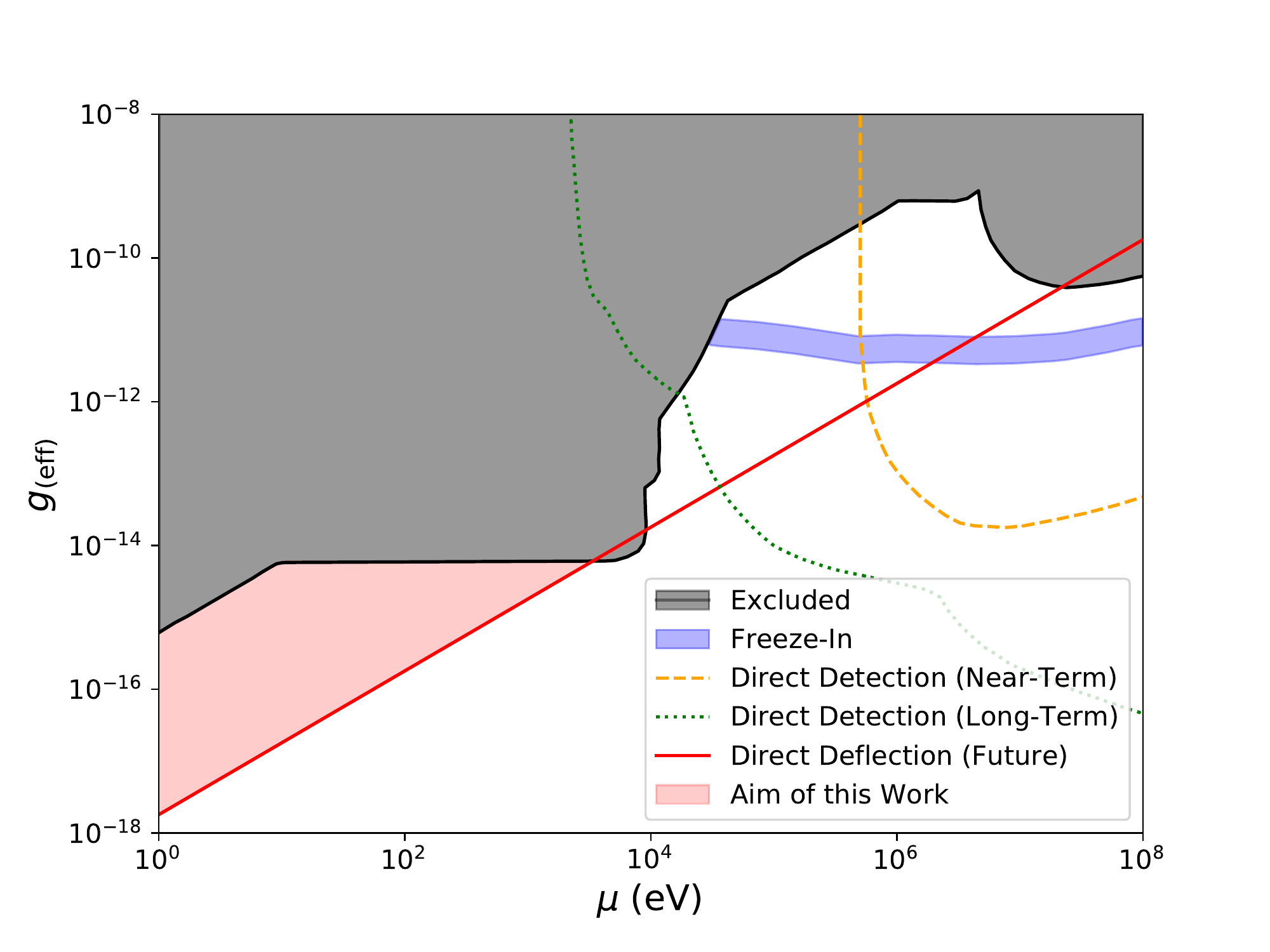}
\caption{A summary of millicharged dark matter parameter space in the $1$ eV to $100$ MeV mass range, showing previously excluded regions as well as the projected sensitivities of future direct deflection \cite{PhysRevLett.124.011801} and direct detection experiments, with the latter loosely divided into near-term (e.g. \cite{SENSEI,SuperCDMS}) and long-term (e.g. \cite{hochberg2021new, 2020Snowmass2021LetterOI}) projects. Also shown is the parameter region motivated by freeze-in \cite{Hall2010,Chu_2012,PhysRevD.99.115009}, and the region that we attempt to motivate in this work.}
\label{fig:motivationPlot}
\end{figure}

A key aspect of any dark matter model is the mechanism for generating its abundance in the universe. Previous models of and searches for millicharged dark matter have focused on millicharges that can be produced through the process of freeze-in, wherein dark matter is slowly produced from rare interactions within the thermal bath of Standard Model particles until the Standard Model temperature falls below the dark matter mass \cite{Hall2010,Chu_2012,PhysRevD.99.115009}. Several previously proposed experiments, in varying stages of development, have the potential to thoroughly search the parameter space of millicharge models compatible with freeze-in \cite{DMRadio,SENSEI,SuperCDMS}. Existing bounds on millicharged dark matter parameters, as well as projected sensitivities of some planned direct detection experiments, are shown in Figure \ref{fig:motivationPlot}.

Freeze-in is not, however, the only means of generating millicharged dark matter, and in fact is only possible in a small part of millicharge parameter space. In particular, dark matter of mass less than a few keV cannot be produced by any thermal mechanism because it would necessarily be hot at keV temperatures, violating bounds on warm dark matter \cite{Irsic_2017, lin2019tasi}. In this paper, we propose a model of ultralight dark matter, viable at significantly lower masses and couplings than those required for freeze-in, wherein the dark matter abundance is the result of an axion-like field misalignment in the early universe, rather than of any thermal mechanism. Notably, a significant region of this light parameter space is detectable using ``direct deflection,'' in which a laboratory electric field separates oppositely-charged dark matter components as they pass through it, leading to a charge separation that can be detected at a downstream detector \cite{PhysRevLett.124.011801}. To the authors' knowledge, this work provides the first cosmologically-consistent theoretical model for a significant region of parameter space to which direct deflection is sensitive. 

We will focus specifically on millicharged dark matter with mass between $1$ eV and $10$ keV. We do not consider higher masses since the lack of stellar cooling constraints above $\mathcal{O}(10\text{ keV})$ means that such masses are already well motivated by freeze-in. Most of the discussion in this work should extend to millicharges with masses below $1$ eV, but some modifications may be required due to the large occupation numbers and de Broglie wavelengths associated with such masses, which are beyond the scope of this work.

This new region of millicharge parameter space also has a rich and distinctive phenomenology. Models of heavy dark matter typically treat dark matter as a particle, and methods of detecting such particle-like dark matter focus on individual scattering processes. Models of light dark matter, on the other hand, typically treat dark matter as a field, and light dark matter detectors correspondingly take advantage of large coherent enhancements at low momentum transfers. Our model bridges the mass region in between these regimes, and has some commonalities with both.  Though our dark
matter behaves as a coherent field at early times and high densities, it does not linearly couple to matter because it carries a conserved charge, which one would typically associate with particle-like models. The most powerful constraints come from the collective impact of scattering processes that are best understood in the particle picture (e.g. stellar cooling via dark matter emission, or friction between dark matter and the primordial baryonic plasma).

We develop our model for true millicharges in Section \ref{sec:modelTheory}, and then modify it to allow for effective millicharges in Section \ref{sec:hiddenPhotons}. We then explore the phenomenology of such ultralight millicharged dark matter, including existing cosmological and experimental constraints on our model and the potential for its detection. We begin with some general cosmological impacts of our model in Section \ref{sec:cosmology}. Then, in Section \ref{sec:pheno}, we walk through a variety of constraints on our model before, finally, discussing the potential for detecting ultralight millicharged dark matter via direct deflection \cite{PhysRevLett.124.011801}.

\section{Model}\label{sec:modelTheory}

To avoid the constraints typically resulting from thermal production mechanisms, our model takes advantage of spontaneous symmetry breaking to generate an initial population of dark matter through an axion-like misalignment mechanism. The simplest symmetry group that we could consider spontaneously breaking is $SU(2)$. In this section, we describe a simple scalar realization of our model, which we use throughout most of this paper. We present an a variant of our model based on quark confinement rather than fundamental scalar fields in Appendix \ref{app:quarkModel}. (An entirely different misalignment-based approach to producing light dark matter has also recently been presented in \cite{ramazanov2021freezein}.)

Since spontaneous symmetry breaking of a field in the fundamental representation of $SU(2)$ leaves no unbroken generators, we consider a field in the adjoint representation of $SU(2)$ or, equivalently, in the fundamental representation of $SO(3)$. We parametrize this field $\phi$ as a triplet of real fields, $(\phi_x,\phi_y,\phi_z)^T$, using $(R_i)_{jk} = -\epsilon_{ijk}$ as the generators of $SO(3)$.

In order for our model to give a light complex field, this $SO(3)$ symmetry will be approximate. However, we will leave $R_3=R_z$ as the generator of an exact, unbroken gauge symmetry which will (for true millicharges) correspond to hypercharge, with gauge field $B_\mu$ and coupling $g/\cos\theta_W$ with $\theta_W$ the Weinberg angle. We will not be concerned with weak interactions in this work, so we will include only electromagnetic couplings below. We note, however, that weak couplings should be expected as well, at least for a typical ultraviolet completion. The most general Lagrangian that we can write down which respects these symmetries is then, up to dimension 4,
\begin{align}\begin{split}
    \mathcal{L} =& -\frac{1}{4}F_{\mu\nu}F^{\mu\nu} + \frac{1}{2}(D_\mu\phi)^T(D^\mu\phi) + \frac{m^2}{2}\phi^T\phi - \frac{\lambda_\phi}{24}(\phi^T\phi)^2 - \frac{\mu_\phi^2}{2}(\phi_x^2+\phi_y^2) \\
    &- \frac{\alpha}{24}\phi^T\phi(\phi_x^2+\phi_y^2) - \frac{\beta}{24}(\phi_x^2+\phi_y^2)^2 \label{eq:LphiFull}
\end{split}\end{align}
where $D_\mu = \partial_\mu + gR_zA_\mu$. Note that the $\mu_\phi^2$, $\alpha$ and $\beta$ terms would be prohibited by an exact $SO(3)$ symmetry, but are expected in the presence of the symmetry-breaking gauge coupling. These terms are naturally generated by radiative corrections; see Appendix \ref{app:EFTofChi}.

\begin{figure}[b]
\centering
\includegraphics[width=0.6\linewidth]{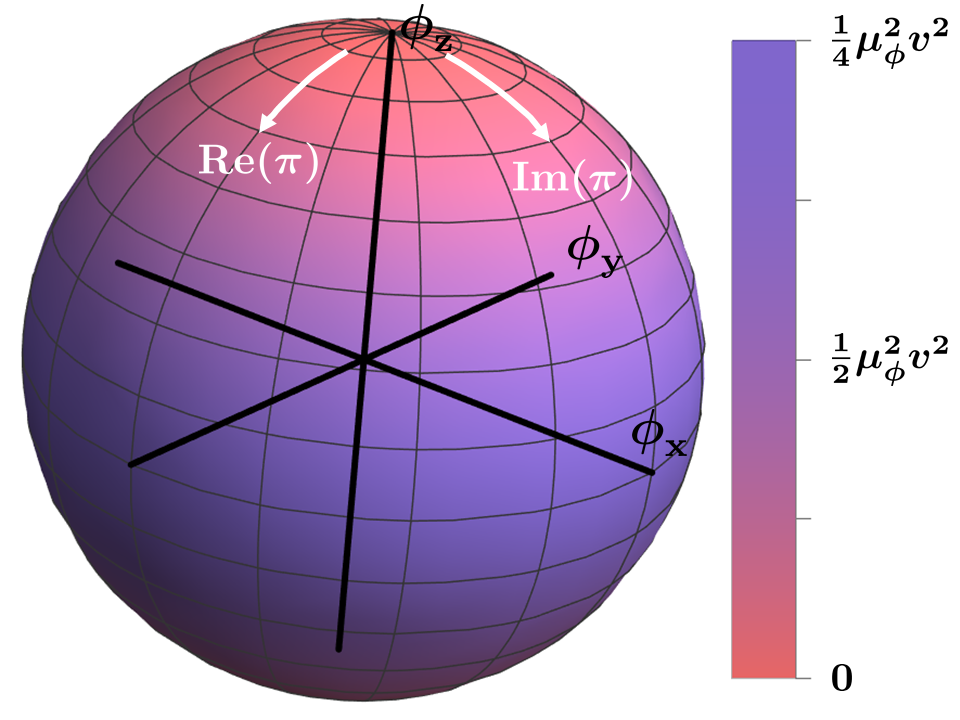}
\caption{The surface of $\phi$ vacua, all with magnitude $|\phi|^2=v^2$, of the $\textrm{SO}(3)$-symmetric theory, colored according to the additional energy density of that field value once the $\mu_\phi^2$ term is included. Also shown are the rotation directions corresponding to the real and imaginary parts of $\chi$.}
\label{fig:spherePlot}
\end{figure}

For $m^2,~\lambda_\phi,~\mu_\phi^2 > 0$, the vacuum of this theory is unique up to an overall sign:
\begin{align}
    \phi_0 &= \pm v\begin{pmatrix} 0 \\ 0 \\ 1 \end{pmatrix} \quad\text{where}\quad v= \sqrt{\frac{6m^2}{\lambda_\phi}}. \label{eq:explicitvev}
\end{align}
These two vacua are related by a $\mathbb{Z}_2$ symmetry. We will assume the positive signed vacuum below, but the physics of both vacua is identical. If we now define $R_\pm = R_x \pm iR_y$, we can expand around this vacuum as
\begin{align}
    \phi = e^{\pi^+R_+ + \pi^-R_-}(1+\sigma)\phi_0
\end{align}
where $\pi^+ = (\pi^-)^* = \pi$ is a complex scalar and $\sigma$ is a real scalar. In the limit of $\mu_\phi\to0$, the $\pi^\pm$ fields are the two Goldstone bosons from breaking the exact $SO(3)$ symmetry. These acquire a mass once the symmetry is made approximate by $\mu_\phi$. This situation is illustrated in Figure \ref{fig:spherePlot}.

We will be interested in the effective field theory at energies far below both $m^2$ and $v^2$, which we assume to be much larger than $\mu_\phi^2$. In this limit, we can ignore the excitations of the mass-$m$ field $\sigma$. (The $\sigma$-dependent terms up to dimension 4 are included in Appendix \ref{app:EFTofChi} for reference.) Then the remaining terms of the Lagrangian are
\begin{align}\begin{split}
    \mathcal{L} \approx& -\frac{1}{4}F_{\mu\nu}F^{\mu\nu} + \frac{v^2}{2}\frac{\sin^2(2|\pi|)}{|\pi|^2} \left|\partial_\mu \pi - ig\pi A_\mu\right|^2 - \frac{v^2}{2}\mu_\phi^2\sin^2(2|\pi|) \\
    &- \frac{v^4}{24}\sin^2(2|\pi|) \left(\alpha + \beta \sin^2(2|\pi|) \right). \label{eq:noSigmaL}
\end{split}\end{align}
Restricting to small excitations of $\pi$, we can expand this up to dimension 4:
\begin{align}\begin{split}
    \mathcal{L} \approx& -\frac{1}{4}F_{\mu\nu}F^{\mu\nu} + 2v^2\left|\partial_\mu \pi - ig\pi A_\mu\right|^2 - v^2\left( 2\mu_\phi^2|\pi|^2 - \frac{8}{3}\mu_\phi^2|\pi|^4 \right) \\
    &- \frac{v^4}{6}\left( \alpha |\pi|^2 - \frac{4\alpha}{3}|\pi|^4 + 4 \beta |\pi|^4 \right) \\
    =& -\frac{1}{4}F_{\mu\nu}F^{\mu\nu} + \left|\partial_\mu \chi - ig\chi A_\mu\right|^2 - \left( \mu_\phi^2 + \frac{v^2\alpha}{12} \right) |\chi|^2 - \left( -\frac{2\mu_\phi^2}{3v^2} -\frac{\alpha}{18} + \frac{\beta}{6} \right) |\chi|^4 \\
    =& -\frac{1}{4}F_{\mu\nu}F^{\mu\nu} + |D\chi|^2 - \mu^2|\chi|^2 - \frac{\lambda}{4}|\chi|^4 \label{eq:chiExpanded}
\end{split}\end{align}
where we have defined the canonically normalized pion field $\chi = \sqrt{2}v\pi$ and the parameters $\mu^2 = \mu_\phi^2 + v^2\alpha/12$ and $\lambda = - 8\mu_\phi^2/(3v^2) - 2\alpha/9 + 2\beta/3$. This leaves us with a low-energy effective field theory containing only two new particles: the light, $U(1)$ charged $\chi$ and $\chi^*$ of mass $\mu$. Note that the full theory is stable even for $\lambda < 0$, as we can see from \eqref{eq:noSigmaL}, though this is not apparent with the terms of higher order in $\chi$ omitted.

We noted earlier that $\mu_\phi^2$, $\alpha$ and $\beta$ receive radiative corrections from loop diagrams; these lead to corresponding contributions to $\mu^2$ and $\lambda$. The precise contributions depend somewhat on one's assumptions about the theory's UV completion: as discussed in Appendix \ref{app:EFTofChi}, a typical UV completion is likely to lead to threshold corrections to the mass, giving
\begin{align}
    \Delta\mu^2 &\sim \frac{g^2\Lambda_{\rm UV}^2}{16\pi^2} \label{eq:loopMuBig} \\
    \Delta\lambda &\sim \frac{g^4 \lambda_\phi}{(16\pi^2)^2}\ln\left(\frac{\Lambda_{\rm UV}^2}{m^2}\right), \label{eq:loopLambda}
\end{align}
where $\Lambda_{\rm UV}$ is the energy scale above which the UV theory applies. In theories whose UV completions give zero threshold corrections, the dominant mass correction instead comes from logarithmic running, which gives
\begin{align}
    \Delta\mu^2 &\sim \frac{g^2\lambda_\phi m^2}{(16\pi^2)^2}\ln\left(\frac{\Lambda_{\rm UV}^2}{m^2}\right) \label{eq:loopMuSmall}.
\end{align}

Note that the one-loop logarithmic contributions to $\mu^2$ and $\lambda$ are both zero. This is most easily understood by working perturbatively around the broken vacuum. The only mass parameter in the broken-vacuum Lagrangian (see Appendix \ref{app:EFTofChi}) comes from the $\sigma$ field, but both $\mu^2$ and $\lambda$ are prohibited in the limit of an exact $SO(3)$ symmetry and thus contributions to them require photon insertions as well. As a result, the leading-order logarithmic contributions come from the gauge field-coupled parts of dimension-5 ($\sigma|D\chi|^2$) and dimension-6 ($\sigma^2|D\chi|^2$) operators, which contribute at two loop order.

\section{Hidden Photon Coupling}\label{sec:hiddenPhotons}

In the model presented above, the new field $\phi$ couples directly to the Standard Model $U(1)_Y$. However, we could also consider a coupling to hidden photons corresponding to a different gauge group $U(1)_D$. This does not change any of the discussion above, except to change all of the $A_\mu$ and $F_{\mu\nu}$ fields to the hidden fields $A_\mu'$ and $F_{\mu\nu}'$. We will consider hidden photons coupled to the Standard Model through a kinetic mixing with the normal photon, such that the Lagrangian corresponding to these two gauge bosons is (for a small mixing parameter $\epsilon\ll1$)
\begin{align}
    \mathcal{L} &= -\frac{1}{4}(F_{\mu\nu}F^{\mu\nu} + F_{\mu\nu}'F^{\mu\nu\prime} + 2\epsilon F_{\mu\nu}F^{\mu\nu\prime}) - \frac{1}{2}m_{A'}^2A_\mu'A^{\mu\prime} + J_\mu^{SM}A^\mu + J_\mu^{DM}A^{\mu\prime}. \label{eq:hiddenPhotonLagrangian}
\end{align}
In this case, interactions between $\chi$ and the Standard Model are mediated by the (massive) hidden photon, while the massless  Standard Model photon does not couple to the dark matter. This ``effective millicharge'' phenomenology is distinct from realizations of true millicharges in models with both kinetic mixing and Stueckelberg mixing among multiple $U(1)$'s \cite{KORS2004366, Cheung_2007, PhysRevD.75.115001}. We will remain agnostic to whether $\chi$ is a true or effective millicharge throughout most of this work, so we assume for simplicity that the hidden photon has a negligible mass. This is not necessary, however, and it is generally straightforward to adapt our results to larger hidden photon masses; we discuss some of the effects of larger hidden photon masses in Section \ref{sec:pheno}.

In the remainder of this paper, we will continue to describe the interaction of the millicharged fields with the Standard Model as if they are true millicharges, without any hidden photons, except when we consider effects that exist only in the presence of hidden photons. As we will see in Section \ref{sec:pheno}, the hidden photon case is in fact more phenomenologically interesting; however, the true millicharge discussion can easily be adapted to the hidden photon case simply by replacing all factors of $g$ with $g_{\rm eff} \coloneqq \epsilon g$ (or more precisely, for $\epsilon \sim 1$, $g_{\rm eff} \coloneqq \epsilon g/\sqrt{1-\epsilon^2}$) when considering interactions with the Standard Model.

A typical way to generate a hidden photon kinetic mixing, assuming it is not present at high energies, is through loops of heavy particles that couple to both $U(1)_Y$ and $U(1)_D$ (see, for example, \cite{PhysRevD.80.015003}). This gives, for $N$ such particle species with $U(1)_Y$ and $U(1)_D$ charges $q_i$ and $Q_i$ respectively, 
\begin{align}
    \epsilon &\sim \frac{g_Yg_D}{16\pi^2}\sum\limits_{i=1}^N q_iQ_i\ln\frac{M^2}{m_i^2}
\end{align}
with $M$ the renormalization scale. Given the known value of $g_Y$ and assuming that $N$, $q_i$, $Q_i$ and $\ln(M^2/m_i^2)$ are $\mathcal{O}(1)$, this mechanism gives $\epsilon \lesssim 10^{-2}g_D$. However, as we will see below, part of the phenomenologically interesting parameter space for a hidden photon-coupled $\chi$ requires $\epsilon \gg g_D$ which is difficult to justify with this loop-level mechanism.

Other methods of generating a kinetic mixing can circumvent this, however \cite{PhysRevD.80.015003}. For example, if the $U(1)_Y$ is embedded in an $SU(5)$ theory at energy scale $M_G$, with a non-zero vacuum expectation value of a Higgs field $\Phi$, the operator $\text{Tr}\left[\Phi F^5_{\mu\nu}\right]F_D^{\mu\nu}$ should generate a kinetic mixing $\epsilon \sim M_G/M_{\rm Pl}$.  Here, $F^5_{\mu\nu}$ and $F^D_{\mu\nu}$ are the $SU(5)$ and $U(1)_D$ field strengths, respectively. This allows for large values of $\epsilon$ even at arbitrarily small couplings $g_D$.

\section{Cosmological Effects}\label{sec:cosmology}

We now consider the cosmological history of a universe in which dark matter consists of the millicharged $\chi$ fields. We begin by summarizing the field evolution of the $\phi$ and $\chi$ fields, including both the initial generation of $\chi$ through misalignment and its late time evolution. Using this history, we work out the conditions for the $\chi$ field to account for all of the dark matter observed today. We also consider possible topological effects on the spontaneous symmetry breaking of $\phi$ to $\chi$. Finally, we consider charge asymmetry of this field, and calculate the photon plasma mass generated by $\chi$. All of the discussion in this section will hold for both true and effective millicharges, subject to the usual $g$ to $g_{\rm eff}$ replacement.

\subsection{Field Evolution}

The $\chi$ field can be cosmologically generated through a misalignment mechanism analogous to that of, for example, the axion \cite{PRESKILL1983127,ABBOTT1983133,DINE1983137,MARSH20161}. At early times when $T \gtrsim \Lambda_{\rm UV} \gtrsim v \sim m$, the $\textrm{SO}(3)$ symmetry is exact and $\phi$ will be thermally distributed with field values much larger than $v$. Once $T$ falls below $\Lambda_{\rm UV}$, a mass difference between the $\phi$ directions is produced by the presence of the gauge coupling, though the theory continues to be well-described by the $\phi$ degrees of freedom if $T \gg v$ since $m^2$ and $\mu_\phi^2$ are negligible at these temperatures. Spontaneous symmetry breaking finally occurs once $T \sim v$, at which point each causally connected patch acquires some initial average value of $\sigma$ and $\pi$ (or equivalently $\chi$). (There will also be some thermal fluctuations, leading to a ``cosmic millicharge background'' of sorts, but this is subdominant and we do not consider it further.) The initial value of $\chi$ should be essentially random: though there is a slight energetic preference for small $|\chi|$, it should have virtually no effect since $\mu \ll v \sim T$ by assumption. The initial value of $|\pi_0| \in [0,\pi/2)$ should then usually be $\mathcal{O}(1)$, absent any mechanism that gives it a preferred value. Then a typical patch will have $|\chi_0| = \sqrt{2}v|\pi_0|$, corresponding to a density of $\rho = 2v^2\mu^2|\pi_0|$. Note that the phases of $\pi_0$ and $\chi_0$ are irrelevant, and in fact we could always redefine the $\phi_x$ and $\phi_y$ directions to make them real.

Immediately after spontaneous symmetry breaking, $\chi$ should evolve according to the usual FRW metric equation of motion:
\begin{align}
    \ddot{\chi} + 3H\dot{\chi} + \mu^2\chi = 0.
\end{align}
We assume throughout this discussion that $\chi$ has negligible spatial dependence, which should be the case for symmetry breaking before the end of inflation; see Section \ref{sub:InflationTopology}. The result of this is that the average value of $\chi$ (i.e. the non-thermal component) should stay constant while $H\gg\mu$. Once $H\ll \mu$, it should instead decrease as a matter field. At these late times, the field should be well-described by a sinusoid of slowly-varying amplitude,
\begin{align}
    \chi(t) \approx \left(\frac{a_0}{a(t)}\right)^{3/2}\sqrt{2}v\pi_0\sin(\mu t) \label{eq:chiSemiclassical}
\end{align}
with $a_0$ the scale factor at which $H=\mu$. We will ignore the transition between these two limits in this work, as it should only give $\mathcal{O}(1)$ corrections.

In fact, as we discuss in Section \ref{sec:pheno}, interactions of $\chi$ with the Standard Model will cause at least part of it to reach higher temperatures, and this may extend to all of $\chi$ depending on the strength of its self-interactions. This does not significantly affect any of the discussion below, however, so we will continue to assume that $\chi$ is dominated by its cold fraction for simplicity.

Assuming a $\Lambda$CDM cosmology, for the values of $\mu$ we consider ($1$ eV to $10$ keV), the transition at $H\sim\mu$ corresponds to a time during radiation domination, giving a present-day average $\chi$ density of
\begin{align}
    \rho_0 \sim 6\times10^{-30}\ \frac{\text{GeV}}{\text{cm}^3}\left(\frac{\mu}{1\ \text{eV}}\right)^{1/2}\left(\frac{v}{1\ \text{GeV}}\right)^2|\pi_0|^2.
\end{align}
We briefly consider the effects of some modified cosmologies below.

Using the most recent Planck measurements of $\rho_{0,\rm{DM}}$ \cite{collaboration2018planck}, this gives a condition on $v$ for $\chi$ to constitute all of dark matter:
\begin{align}
    v \sim 5\times10^{11}\ \text{GeV}\ \left(\frac{\mu}{1\ \text{eV}}\right)^{-1/4}|\pi_0|^{-1}. \label{eq:vMuRelation}
\end{align}

We noted in Section \ref{sec:modelTheory} that radiative corrections lead to a minimal natural scale for $\mu^2$. In the presence of threshold corrections, i.e. \eqref{eq:loopMuBig}, this minimal value motivates
\begin{align}
    g \sim 6\times10^{-20}\left(\frac{\mu}{1\text{ eV}}\right)^{5/4}\left(\frac{\Lambda_{\rm UV}}{m}\right)^{-1}\frac{|\pi_0|}{\sqrt{\lambda_\phi}}. \label{eq:radiativeMotivationBig}
\end{align}
As we will see in Section \ref{sec:pheno}, both existing and planned experiments are only sensitive to couplings of $g\gtrsim 10^{-18}(\mu/1\text{ eV})$. This suggests that models testable in the near term would have masses $\mu^2$ fine-tuned around the $10^{-3}$ level, assuming both $\lambda_\phi$ and $|\pi_0|$ are order unity. Note that larger values of $g$ at a given $\mu$ can be consistent with smaller values of $\lambda_\phi$, however, without the need for fine tuning.

In UV theories with zero threshold corrections, i.e. \eqref{eq:loopMuSmall}, the motivated coupling is instead
\begin{align}
    g \sim 8\times10^{-19}\left(\frac{\mu}{1\text{ eV}}\right)^{5/4}\frac{|\pi_0|}{\lambda_\phi\sqrt{\ln(\Lambda_{\rm UV}^2/m^2)}}. \label{eq:radiativeMotivationSmall}
\end{align}
In this case, parameter space testable in the near future appears without any fine tuning of $\mu^2$ for any perturbative $\phi^4$ vertex, i.e. for $\lambda_\phi \lesssim 1$.

$\Lambda$CDM is not the only allowed model of the early universe, however, and the relationship between $\mu$, $g$ and $\lambda$ changes in many modified cosmologies. Here we consider the effects of two examples of such modifications, with opposite effects: a period of early matter domination and a period of early kination. In both cases the effects on our model precisely mirror those on axion-like particles \cite{PhysRevD.100.015049}. 

Matter domination at the time of spontaneous symmetry breaking motivates larger initial values of $\chi$, as the more rapid expansion of the universe during this period tends to dilute the millicharge density compared to the radiation-dominated case. This change is maximal if the universe is matter dominated from the entire period from $H\sim\mu$ to reheating, with a low reheating temperature ($\sim10$ MeV). This can reduce the natural scale of $g$ at a given $\mu$ by a factor of up to approximately $(10\text{ MeV}/\mu)^{1/2}$, relative to \eqref{eq:radiativeMotivationBig} or \eqref{eq:radiativeMotivationSmall}.

An early period of kination, in which the energy density of the universe is dominated by the kinetic energy of a scalar field, has the opposite effect, due to the rapid dilution of this kinetic energy. In this case, the radiatively-motivated value of $g$ can instead increase by roughly the same factor, $(10\text{ MeV}/\mu)^{1/2}$.

Other modifications to $\Lambda$CDM may have even larger effects on radiatively motivated couplings, but we will not attempt to exhaustively catalogue these here; our key point is that \eqref{eq:radiativeMotivationBig} and \eqref{eq:radiativeMotivationSmall} should be understood to have a few orders of magnitude of uncertainty depending on the cosmology of the early universe, especially for smaller values of $\mu$.

\subsection{Inflation and Topology}\label{sub:InflationTopology}

In the discussion above, we implicitly assumed that spontaneous symmetry breaking of $\phi$ to $\chi$ leads to a uniform initial value of $\chi$. In actuality, casually disconnected patches of the universe will have independent initial misalignments.

Again mirroring the axion scenario, the present-day effects of these patch differences depend heavily on whether spontaneous symmetry breaking occurs during or after inflation \cite{MARSH20161}. Spontaneous symmetry breaking occurs after inflation if (up to $\mathcal{O}(1)$ factors) $v < T_I = H_I/2\pi$, where $H_I$ is the inflationary Hubble scale and $T_I$ is the corresponding Gibbons-Hawking temperature. In this case, when $T \sim v$, different causally disconnected patches reach different initial values of $\chi$. Since this occurs after the end of inflation, this leads to many patches with different initial values of $\chi$ within the present observable universe.

Symmetry breaking after inflation also leads to a second effect, namely topological defects. In fact, two different types of defects are present: first, as we noted above \eqref{eq:explicitvev}, there are two minima of the potential of $\phi$, so causally disconnected patches can reach different vacuum expectation values $\pm v$ of $\phi$. Second, because the second homotopy group $\pi_2(S^2) = \mathbb{Z}$, it is possible for individual patches to have topologically nontrivial configurations of the field $\chi$ (recall that $\chi$ and $\chi^*$ together specify a position on the sphere of approximate minima of the potential of $\phi$).

Domain walls between distinct different signs of $v$ are likely to be inconsistent with observed cosmology, as domain wall energy density scales as $a^{-2}$ and therefore eventually dominates the energy of the universe for most models \cite{PhysRevLett.48.1156}. There may be additional phenomenology associated with the decay of higher-energy configurations of $\chi$ within patches, but we do not consider this further as the domain wall problems alone make this scenario strongly disfavored.

The cosmologically preferred scenario is therefore symmetry breaking before the end of inflation. In this case, the current observable universe originated within a single causally connected patch at the time of symmetry breaking, giving a uniform initial value of $\chi$ within it. While topologically nontrivial configurations can still occur, their corresponding energy densities are dramatically reduced by subsequent inflation, so they should not be cosmologically catastrophic.

Symmetry breaking of $\phi$ to $\chi$ before the end of inflation does, however, lead to isocurvature perturbations (see, for example, \cite{MARSH20161} for a summary of the analogous axion case). Observations of the cosmic microwave background constrain the amplitude of isocurvature perturbations, which impose a limit on the Hubble scale at inflation $H_I$ as a function of the symmetry breaking scale $v$,
\begin{align}
    \frac{H_I}{v} \lesssim 4\times10^{-5}|\pi_0|,
\end{align}
or, equivalently, in terms of the mass $\mu$,
\begin{align}
    H_I \lesssim 2\times10^7\text{ GeV}\left(\frac{\mu}{1\text{ eV}}\right)^{-1/4}|\pi_0|.
\end{align}
Since the Hubble scale at inflation is not currently known, this does not impose any additional constraints on our model. A future measurement of $H_I$ would place an upper bound on $\mu$, however. Conversely, assuming that $\chi$ is in the mass range we consider precludes models of inflation with Hubble scale above $3\times10^{6-7}$ GeV. Our model thus favors models of low-scale inflation; see, for example \cite{LowScaleInflationGerman, LowScaleInflationTakahashi, LowScaleInflationChoudhury}.

\subsection{Plasma Mass}\label{sec:plasmaMass}

Since $\chi$ is electromagnetically coupled, a background $\chi$ field will give a plasma mass to the photon (or, in the hidden photon case discussed below, an additional mass to the hidden photon). We are not concerned with relativistic effects, so this is given by the usual, classical result $m_{\rm plasma}^2 = nq^2/m$, with $n$ the number density of charged particles, $q$ their charge, and $m$ their mass \cite{PhysRev.33.195}. Writing this mass contribution in terms of the number densities of positively (negatively) charged $\chi$ particles $n_+$ ($n_-$), we have
\begin{align}
    \Delta m_\gamma^2 &= \frac{(n_+ + n_-)g^2}{\mu}.
\end{align}

For a spatially uniform field ($\partial_i\chi = 0$) with no background electromagnetic fields (so that, in a convenient gauge, $A_\mu=0$), the density and charge density of $\chi$ are given by
\begin{subequations}\begin{align}
    \rho_\chi &= \mu(n_+ + n_-) \\
    J_0 &= g(n_+ - n_-).
\end{align}\end{subequations}
But we can also see from \eqref{eq:chiExpanded} that
\begin{subequations}\begin{align}
    T_{00} = \rho_\chi &= \mu^2\chi\chi^* + \dot{\chi}\dot{\chi}^* \label{eq:T00chi} \\
    J_0 &= ig(\chi^*\dot{\chi} - \chi\dot{\chi}^*). \label{eq:J0chi}
\end{align}\end{subequations}
Then
\begin{align}
    \Delta m_\gamma^2 &= \frac{g^2}{\mu^2}\rho_\chi.
\end{align}

For the local dark matter density of $0.3$ GeV/cm$^3$, and assuming other contributions to the mass are negligible, this gives a local (dark) photon plasma mass of
\begin{align}
    m_\gamma &\approx 2\times10^{-18}\text{ eV}\left(\frac{g}{10^{-15}}\right)\left(\frac{\mu}{1\text{ eV}}\right)^{-1}. \label{eq:DPplasmamass}
\end{align}

A more formal derivation of this result, which explicitly accounts for the background field being coherent, is presented in \cite{PhysRevA.94.012124}. More extensive discussion of the mode structure of scalar QED plasmas can also be found in that work.

\subsection{Charge Asymmetry}

We can see from \eqref{eq:chiExpanded} that the current associated with the field $\chi$ alone is given by
\begin{align}
    ig(\chi^*\partial^\mu\chi - \chi\partial^\mu\chi^*) &= 2g\ \text{Im} (\chi\partial^\mu\chi^*).
\end{align}
This allows for a potential net charge asymmetry of the universe after spontaneous symmetry breaking, which is heavily constrained by the small measured anisotropy of the cosmic microwave background \cite{Caprini_2005}. Assuming a uniform distribution of dark matter, this bound is
\begin{align}
    |\rho_0| &\leq 2\times10^{-74}\ h^2\ \text{GeV}^3.
\end{align}
Once $H \lesssim \mu$, the charge density of $\chi$ evolves in the same way as its mass density (discussed above), so an initial value of $|\rho| = g|\chi^*\partial^\mu\chi - \chi\partial\chi^*|$ will correspond to a present-day charge density of
\begin{align}
    |\rho_0| &\approx 2\times10^{-53}\left(\frac{\mu}{1\ \text{eV}}\right)^{-3/2}g|\chi^*\partial^\mu\chi - \chi\partial^\mu\chi^*|_{H=\mu}
\end{align}
with $\chi$ evaluated when $H=\mu$. This gives a bound of
\begin{align}
    g|\chi^*\partial^\mu\chi - \chi\partial^\mu\chi^*|_{H=\mu} &\leq 5\times10^5\ \text{eV}^3\left(\frac{\mu}{1\ \text{eV}}\right)^{3/2}
\end{align}
or, in dimensionless form
\begin{align}
    g\left|\frac{\chi^*\partial^\mu\chi - \chi\partial^\mu\chi^*}{\mu\chi^2}\right|_{H=\mu} &\lesssim 1\times10^{-36}\left(\frac{\mu}{1\ \text{eV}}\right),
\end{align}
an extremely tight constraint.

Note, however, that the charge asymmetry of $\chi$ is dramatically diluted by inflation: since the value of the $\chi$ field is essentially constant while $H \gg \mu$, the derivative coupling makes the charge density of $\chi$ dilute as the first power of the scale factor, $\rho \propto a^{-1}$. Thus comparatively large charge asymmetries prior to inflation are consistent with observations.

Moreover, so long as the relevant $U(1)$ symmetry ($U(1)_Y$ or, for hidden photons, $U(1)_D$) is exact at the time that spontaneous symmetry breaking of the $\phi$ field occurs, charge conservation should be preserved during the symmetry breaking process. Then if there was no net charge associated with the $\phi$ field before symmetry breaking, there should be no net charge associated with $\chi$ afterwards, i.e. $\int dV\ \text{Im} (\chi\partial^\mu\chi^*) = 0$. UV completions in which the $U(1)$ symmetry is \textit{not} exact, however, may be constrained by these present-day charge asymmetry observations.

Furthermore, a charge asymmetry does not have to be uniform or global to lead to observable signatures. Bounds have also been derived for stochastic charge distributions, for which the charge within any chosen volume has zero expectation value but nonzero variance \cite{Caprini_2005}. Such a situation occurs in our model even if there is zero total charge asymmetry, due to the nature of coherent states of charged fields.

To see this, consider the quantum mechanical description of the field state of $\chi$, which we had previously described classically by \eqref{eq:chiSemiclassical}. Once $H \ll \mu$, we expect $\chi$ to be well-described by a spatially uniform coherent state. Let $\hat{a}_\mathbf{k}^\dagger$ and $\hat{b}_\mathbf{k}^\dagger$ be the creation operators at momentum $\mathbf{k}$ for $\chi$ and $\chi^*$ respectively and $\Pi$ be the canonical conjugate field to $\chi$, so that
\begin{align}\begin{split}
    \chi(\mathbf{x}) &= \int \frac{d^3\mathbf{p}}{(2\pi)^3} \frac{1}{\sqrt{2\omega_\mathbf{p}}} \left(\hat{a}_\mathbf{p}e^{i\mathbf{p}\cdot\mathbf{x}} + \hat{b}_\mathbf{p}^\dagger e^{-i\mathbf{p}\cdot\mathbf{x}}\right) \\
    \Pi(\mathbf{x}) &= i \int \frac{d^3\mathbf{p}}{(2\pi)^3} \sqrt{\frac{\omega_\mathbf{p}}{2}} \left(\hat{a}_\mathbf{p}^\dagger e^{i\mathbf{p}\cdot\mathbf{x}} - \hat{b}_\mathbf{p} e^{-i\mathbf{p}\cdot\mathbf{x}}\right).
\end{split}\end{align}
Then the post-symmetry breaking state can be obtained from the usual displacement operator, giving
\begin{align}\begin{split}
    \left|\chi_0\right> &= e^{-i\int d^3\mathbf{x} \sqrt{\mu} \left(\chi_0\Pi(\mathbf{x})+\chi_0^*\Pi^*(\mathbf{x})\right)}\left|0\right> \\
    &= e^{\frac{1}{\sqrt{2}}\left(\chi_0\hat{a}^\dagger+\chi_0^*\hat{b}^\dagger\right)-\frac{1}{\sqrt{2}}\left(\chi_0^*\hat{a}+\chi_0\hat{b}\right)}\left|0\right> \\
    &= e^{-|\chi_0|^2/2}\sum\limits_n \frac{1}{ n!}\left(\frac{\chi_0\hat{a}^\dagger+\chi_0^*\hat{b}^\dagger}{\sqrt{2}}\right)^n\left|0\right>,
\end{split}\end{align}
where we have suppressed the momentum subscripts, which are zero throughout. This coherent state has a charge variance per mode (i.e. $\left\langle g^2(a^\dagger a - b^\dagger b)^2\right\rangle$) of $g^2|\chi_0|^2$. Since $|\chi_0|^2$ is the expected number of particles $N$ in the mode (including both charges), this gives a variance in the total charge of a given mode $\mathbf{k}$ of $\left\langle q^2 \right\rangle_\mathbf{k} = g^2 N_\mathbf{k}$. So long as the variation in different modes is independent, this means that \textit{any} coherent collection of $N$ $\chi$ particles will have a root-mean-square charge of $g\sqrt{N}$.

The root-mean-square charge density of the universe is constrained on lengthscales of order kilo- to megaparsecs by various cosmological observations \cite{Caprini_2005, Rudnick:2003qf, PhysRevLett.85.700, PhysRevD.61.043001, PhysRevD.65.023517}. The exact bound resulting from these observations is dependent on the spectral indices of the resulting magnetic fields and vorticities, but rough estimates give
\begin{align}
    g \lesssim 3\times10^{-(1-5)}\left(\frac{\mu}{1\text{ eV}}\right)^{1/2}.
\end{align}
This is far weaker than various constraints on $g$ that we derive below, so we do not attempt to refine this estimate further.

\section{Phenomenology}\label{sec:pheno}

We now turn to the phenomenology of generic light millicharged dark matter, exploring specific effects of such millicharges and the constraints that those effects allow us to derive. While the cosmological history described in the previous section is mostly specific to our model of $\chi$, the discussion in this section should extend to most models of light millicharges with minimal modifications, so long as the dark matter is produced much colder than the Standard Model bath. We will continue to refer to our model for concreteness. 

Since some of the constraints in this section depend on whether the millicharges couple directly to the Standard Model photon or through a kinetically mixed dark photon, this section is organized as follows: We begin by considering constraints that affect $\chi$ independently of kinetic mixing. We then consider constraints that affect only true millicharges that couple directly to the photon, followed by constraints that affect only effective millicharges that couple to a hidden photon. Finally, we discuss an upcoming experiment which should be able to search for millicharges in a significant section of the parameter space we consider in this work.

These constraints are summarized in Figures \ref{fig:MillichargeConstraints}, \ref{fig:DPConstraints} and \ref{fig:maxgeff}. Figure \ref{fig:MillichargeConstraints} shows several of the bounds on true millicharges which couple directly to the Standard Model $U(1)_Y$, along with the projected reach of direct deflection experiments. Figure \ref{fig:DPConstraints} shows constraints and projected reaches in the $(g,g_{\rm eff})$ plane for effective millicharges with various masses $\mu$. Finally, Figure \ref{fig:maxgeff} shows the maximum value of $g_{\rm eff}$ as a function of $\mu$ consistent with existing bounds on effective millicharges, which can be compared to the direct deflection sensitivity projections.

\subsection{Existing Constraints}

\subsubsection{Constraints on Both True and Effective Millicharges}

There are a few constraints on the charge and mass of $\chi$ as a result of its thermal history. First, while its non-thermal production mechanism ensures that $\chi$ is initially cold, it must remain non-relativistic at a photon temperature of $T_\gamma \sim$ keV in order to satisfy the \textbf{warm dark matter bound} (see, for example, \cite{Irsic_2017, lin2019tasi}). Matter power spectrum modes that enter the Hubble horizon while dark matter is relativistic can be washed out by rapidly moving dark matter. This leads to a suppression of density perturbations at small scales, which can be constrained with observations of, for example, the Lyman-$\alpha$ forest.

To calculate the thermal evolution of the $\chi$ field, we consider its scattering off of generic charged particles. For a Coulomb-like interaction, the tree-level momentum-transfer cross-section in vacuum is formally IR divergent. This divergence is regulated by the Debye screening length $\lambda_D$. This screening results in scattering of $\chi$ from the Standard Model being dominated by rare but high-energy scattering events, which excite a small fraction of the $\chi$ abundance to energies close to the Standard Model temperature $T$.  We may then consider two limiting cases: if the characteristic timescale for $\chi$ self-scattering (controlled by $\lambda$ and $g$, not $g_{\rm eff}$) is less than $H^{-1}$, then the $\chi$ population thermalizes to a common temperature $T_\chi$ (distinct from $T$), and warm dark matter bounds are applicable.  Alternatively, if the timescale for $\chi$ self-scattering exceeds $H^{-1}$, then the dominant $\chi$ population remains cold despite the accumulation of a smaller population of energetic $\chi$. We will consider each case in turn, devoting more attention to the case of $\chi$ thermalization since it is the one constrained by warm dark matter bounds.

In the regime of interest, when $gq_f \ll \lambda_D \mu \beta^2$ (with $q_f$ the electric charge of the other particle and $\beta$ its velocity), the total center-of-mass momentum-transfer cross section for $\chi$ scattering off a charged particle $f$ has been calculated to be \cite{PhysRevD.100.123011, PhysRevLett.104.151301, PhysRevD.87.115007}
\begin{align}
    \sigma_{\rm MT}(\chi f\to\chi f) \approx \frac{g^2q_f^2}{4\pi \beta^2\mathbf{k}^2}\ln\left(\frac{4\pi\lambda_D\mu \beta^2}{gq_f}\right)
\end{align}
with $v$ the relative velocity in the center-of-mass frame, and $\mathbf{k}$ the momentum of either particle in that frame. Since the number density of Standard Model particles is exponentially suppressed for $T < m_f$, we can consider only contributions from relativistic species and set $\beta=1$. For relativistic $f$, the center-of-mass momentum is approximately
\begin{align}
    |\mathbf{k}_{\rm CM}| \sim \frac{(\mu+2T_\chi)T}{\sqrt{m_f^2 + 2(\mu+2T_\chi)T}},
\end{align}
where $T_\chi\ll T$ is the temperature (or, equivalently, the momentum scale) of $\chi$ before scattering. Then, up to an $\mathcal{O}(1)$ factor,
\begin{align}
    \langle\sigma_{\rm MT} \beta\rangle_{\chi f\to\chi f} \sim \frac{g^2q_f^2}{(\mu+2T_\chi)T}\left(\frac{m_f^2+2(\mu+2T_\chi)T}{(\mu+2T_\chi)T}\right).
\end{align}
Following such a scattering event, $\chi$ acquires an energy well approximated by
\begin{align}
    \Delta T_\chi \sim (T-T_\chi)\left(\frac{(\mu+2T_\chi)T}{m_f^2 + 2(\mu+2T_\chi)T}\right)(1-\cos\theta)
\end{align}
with $\theta$ the scattering angle. Note that this has the same angular dependence as the momentum-transfer cross section. Define
\begin{align}
    \Delta T_\chi^{\rm MT} = (T-T_\chi)\left(\frac{(\mu+2T_\chi)T}{m_f^2 + 2(\mu+2T_\chi)T}\right).
\end{align}
Then the temperature evolution of $\chi$ is given by
\begin{align}\begin{split}
    \frac{\partial T_\chi}{\partial \ln T} &\sim T_\chi - \sum\limits_{f;m_f\lesssim T} \frac{n_f\langle\sigma_{\rm MT} \beta\rangle_{\chi f\to\chi f}}{H(T)}\Delta T_\chi^{\rm MT}(T, T_\chi) \\
    &\sim T_\chi - \sum\limits_{f;m_f\lesssim T} g^2q_f^2\frac{m_{\rm Planck}}{\mu+2T_\chi}(T-T_\chi) \label{eq:Tchievolution}
\end{split}\end{align}
where the first term accounts for redshift  and the sum is over relativistic Standard Model degrees of freedom. The initial condition for this evolution is given by $T_\chi \sim 0$ at the end of inflation; the warm dark matter bound requires that $T_\chi \lesssim \mu$ once $T \sim 1$ keV.

For $\mu \ll T_\chi \ll T$, \eqref{eq:Tchievolution} gives $\partial T_\chi/\partial \ln T=0$ when
\begin{align}
    T_\chi = T_\chi^* &\sim gq_{\rm rms}\sqrt{m_{\rm Planck}TN_f}
\end{align}
where $N_f$ is the number of relativistic degrees of freedom at temperature $T$ and $q_{\rm rms}$ is their root-mean-square electric charge. For a given set of relativistic particles, the $T/(\mu+2T_\chi)$ term in \eqref{eq:Tchievolution} causes $T_\chi$ to rapidly asymptotically approach $T_\chi^*$; when the population of a given degree of freedom plummets as the universe cools below its mass, $T_\chi$ consequently quickly approaches the new asymptote. This continues until the last charged Standard Model particles---the electron---becomes non-relativistic. After this point, $T_\chi$ redshifts as radiation. The warm dark matter bound therefore reduces to confirming that
\begin{align}
    T_\chi^*(m_e)\times\frac{1\text{ keV}}{m_e} \lesssim \mu
\end{align}
which gives us
\begin{align}
    g \lesssim 1\times10^{-14} \left(\frac{\mu}{1\text{ eV}}\right). \label{eq:WDMbound}
\end{align}
This estimate was confirmed with good precision using numerical simulation of \eqref{eq:Tchievolution}.

We noted earlier that $\chi$ may not thermalize for sufficiently small couplings $g$ and $\lambda$. In this case, only a small fraction $\sum n_f\langle\sigma_{\rm MT} \beta\rangle/H \lesssim 10^{-6}$ of $\chi$ particles are heated, while the rest remain cold at all times. The small number of hot particles act essentially as additional neutrinos, but their effects are not constrained by current observations because they make up a small fraction of the universe's energy budget while being heavier and colder than Standard Model neutrinos. We can thus use \eqref{eq:WDMbound} as a conservative bound on $\chi$.

This scattering of $\chi$ from charged Standard Model particles---specifically protons---also leads to another constraint, due to the effect it has on the Standard Model plasma after recombination. The main effect of \textbf{dark matter-proton scattering} on cosmological observables comes from a modification of baryon acoustic oscillation amplitudes \cite{PhysRevD.98.123506, PhysRevD.98.103529}. This is caused both by frictive damping of modes after they enter the cosmological horizon and through a modification of the Sachs-Wolfe effect \cite{Buen_Abad_2018}. Other effects of dark matter-proton scattering are described in \cite{PhysRevD.98.123506}. Using CMB temperature and polarization power spectra, these effects impose a bound on the dark matter-proton scattering cross section.

For the low masses ($\mu \ll$ GeV) that we are considering, these effects give an essentially constant bound on the dark matter-proton scattering momentum transfer cross section, $\left\langle\sigma_{\rm MT} \beta^4\right\rangle_{\chi p\to\chi p} < 2 \times 10^{-41}$ cm$^2$. Since essentially all protons are nonrelativistic after recombination, the total momentum-transfer cross section is
\begin{align}
    \left\langle\sigma_{\rm MT} \beta^4\right\rangle_{\chi p\to\chi p} \sim \frac{g^2e^2}{4\pi\mu^2}\ln\left(\frac{4\pi\lambda_D\mu \beta^2}{ge}\right).
\end{align}
Using values from the time of recombination, the resulting bound on $g/\mu$ is roughly
\begin{align}
    g \lesssim 1\times10^{-15}\left(\frac{\mu}{1\text{ eV}}\right),
\end{align}
slightly stronger than the warm dark matter bound. Testing the effects of dark matter-proton scattering on halo formation based on Milky Way satellite observations gives a similar constraint \cite{Nadler_2019}, but the analysis of that work would need to be repeated for our model in order to obtain an exact bound.

Finally, we can similarly confirm that $\chi$ annihilation in the early universe would not significantly impact CMB anisotropies. In particular, \textbf{energy injection from dark matter annihilation} during the cosmic dark ages leads to additional ionization, thereby altering the anisotropies of the cosmic microwave background \cite{PhysRevD.93.023527}. CMB measurements therefore lead to a bound on the dark matter annihilation cross section of roughly $\langle\sigma \beta\rangle_{\rm ann} \lesssim 10^{-36}(\mu/1\text{ eV})$ cm$^3$/s .

Since the $\chi$ field is too light and too cold to significantly annihilate to any charged Standard Model particles, the only relevant annihilation process is $\chi\chi^* \to \gamma\gamma$. It is easy to see that all three tree-level contributions to this process will give matrix elements of order $g^2$ for cold $\chi$, and therefore scattering cross-sections of order $g^4/\mu^2$ (times a logarithmic factor). Then CMB anisotropy measurements require
\begin{align}
    g \lesssim 2\times10^{-9}\left(\frac{\mu}{1\text{ eV}}\right)^{3/4}
\end{align}
which is far weaker than the warm dark matter and dark matter-proton scattering bounds we derived previously for the masses we are interested in ($1\text{ eV} \lesssim \mu \lesssim 10^4\text{ eV}$).

There are a few other constraints that apply to both true and effective millicharges. First, as discussed above, millicharged dark matter leads to a plasma mass for photons within it. The \textbf{mass of the photon} has been constrained in a variety of ways, and we will not attempt to describe them all here; see, for example, the table of limits in \cite{PDGphoton}. Many of these limits, however, are specific to a Proca mass, and do not apply to the plasma mass resulting from $\chi$. In particular, present-day astrophysical analyses generally cannot constraint the plasma mass from $\chi$ significantly below the plasma mass resulting from local interstellar free electrons, which give a plasma mass of roughly $10^{-14}$ eV even in intergalactic space \cite{CONLON2018169}. 

Accounting for this, the strongest constraint on $\chi$ actually comes from Earth-based tests of Coulomb's law \cite{PhysRevLett.26.721}, which constrain $m_\gamma \lesssim 10^{-14}$ eV and apply equally well to Proca and plasma masses. This corresponds to a bound of
\begin{align}
    g \lesssim 7\times10^{-12}\left(\frac{\mu}{1\text{ eV}}\right).
\end{align}
(In the case of effective millicharges, one needs to be careful about the basis used to describe normal and dark photons: though one mass eigenstate remains massless in this case, this mass eigenstate is not what couples to Standard Model charges. The bound from Coulomb's law tests therefore still applies.) This is a fairly weak bound, and in fact is not even the strongest bound on $\chi$ from this experiment: interpreting the apparatus of \cite{PhysRevLett.26.721} as a direct deflection experiment instead, as was done in \cite{PhysRevLett.124.011801}, gives a considerably stronger bound of approximately
\begin{align}
    g \lesssim 2\times10^{-13}\left(\frac{\mu}{1\text{ eV}}\right).
\end{align}
We use this stronger bound in the figures below.

It is worth noting that, while this photon mass bound is weaker than some of the other bounds on millicharges in the mass range we consider (for example the dark matter-proton scattering bound), it remains the most sensitive existing laboratory probe for them. It may, however, soon be surpassed in this regard by direct deflection experiments, discussed in Section \ref{sec:detection}. Higher-mass millicharges may also be detectable using direct detection in condensed matter systems (see, for example, \cite{DD_KahnLin, DD_Essig}).

An additional bound arises from constraints on \textbf{stellar cooling} due to $\chi^\pm$ pair emission in stars. Stellar interiors can produce $\chi,\chi^*$ pairs, which escape the star due to their weak coupling. This leads to additional cooling of the star, which can be constrained by astrophysical observations. The strongest bound for our masses of interest comes from observations of red giants \cite{Vogel_2014}. For masses $\mu$ between $1$ eV and $10$ keV, the corresponding bound is essentially constant: 
\begin{align}
    g \lesssim 6\times10^{-15},
\end{align}
which is strong enough to render the Coulomb's law test bound irrelevant throughout our mass range. 

Analogous bounds from the \textbf{cooling of the 1987A supernova} are weaker for all millicharge masses that we consider \cite{Chang2018}, though at higher masses they become significant as the stellar cooling bounds no longer apply.

Finally, \textbf{dark matter self-interactions} are also constrained by observations of the shapes of dark matter halos. In particular, cosmological simulations indicate that self-interactions lead to dark matter densities near the centers of dwarf galaxy halos that are below their observed values \cite{PhysRevD.96.115021, TULIN20181}. This leads to a bound on the $\chi\chi$ scattering cross section over $\mu$. For effective millicharges, this cross section depends on the hidden photon mass, which we have not specified so far. Restricting to Standard Model photons or light hidden photons ($m_\gamma' \ll \mu \beta_{\rm DM}$), however, this bound is approximately
\begin{align}
    g \lesssim 1\times10^{-8}\left(\frac{\mu}{1\ \text{eV}}\right)^{3/4}.
\end{align}
Note that, since this bound does not depend on any coupling to the Standard Model, it is a bound on $g$ rather than $g_{\rm eff}$ for effective millicharges. 

The dark matter self-interaction bound also places a constraint on $\lambda$:
\begin{align}
    \lambda \lesssim 1\times10^{-10} \left(\frac{\mu}{1\text{ eV}}\right)^{3}.
\end{align}
This upper bound is far larger than the radiatively generated contribution to $\lambda$, given by \eqref{eq:loopLambda}, for all allowed couplings $g$ within our chosen mass range (assuming the $\phi^4$ coupling is perturbative, i.e. $\lambda_\phi \lesssim 1$).

\subsubsection{Constraints on True Millicharges Only}

There is one bound specific to true millicharges, which comes from \textbf{deceleration of rotating galaxies}. Millicharged particles in the halo of a rotating galaxy are deflected by the galaxy's magnetic field as they pass through the galactic disk, leading to a net deceleration of the galaxy's rotation. Demanding that the timescale of this deceleration be less than a galactic lifetime requires, very roughly \cite{Stebbins_2019}, 
\begin{align}
    g \lesssim 10^{-22}\left(\frac{\mu}{1\text{ eV}}\right).
\end{align}
Regardless of the uncertainty in this bound, however, it is unambiguously the strongest constraint on true millicharge for all masses we consider.

In principle, analogous bounds for effective millicharges can be derived as well. However, the non-zero mass of the dark photon effectively limits the range of the coherent magnetic fields responsible for galactic spin-down, exponentially suppressing fields with coherence length greater than the inverse dark photon mass. Galactic magnetic fields typically have coherence lengths on the order of kiloparsecs \cite{galacticMagneticFields}, corresponding to a photon mass of roughly $10^{-(26-27)}$ eV. Galactic spin-down bounds should therefore be exponentially weakened for dark photon masses above this threshold, though a thorough analysis of this case has not yet been conducted. Given this uncertainly and our general agnosticism towards the dark photons mass, we do not include a galactic spin-down bound in our effective millicharge constraints, though we note that one may be derived in the future.

\begin{figure}[b]
\centering
\includegraphics[width=0.7\linewidth]{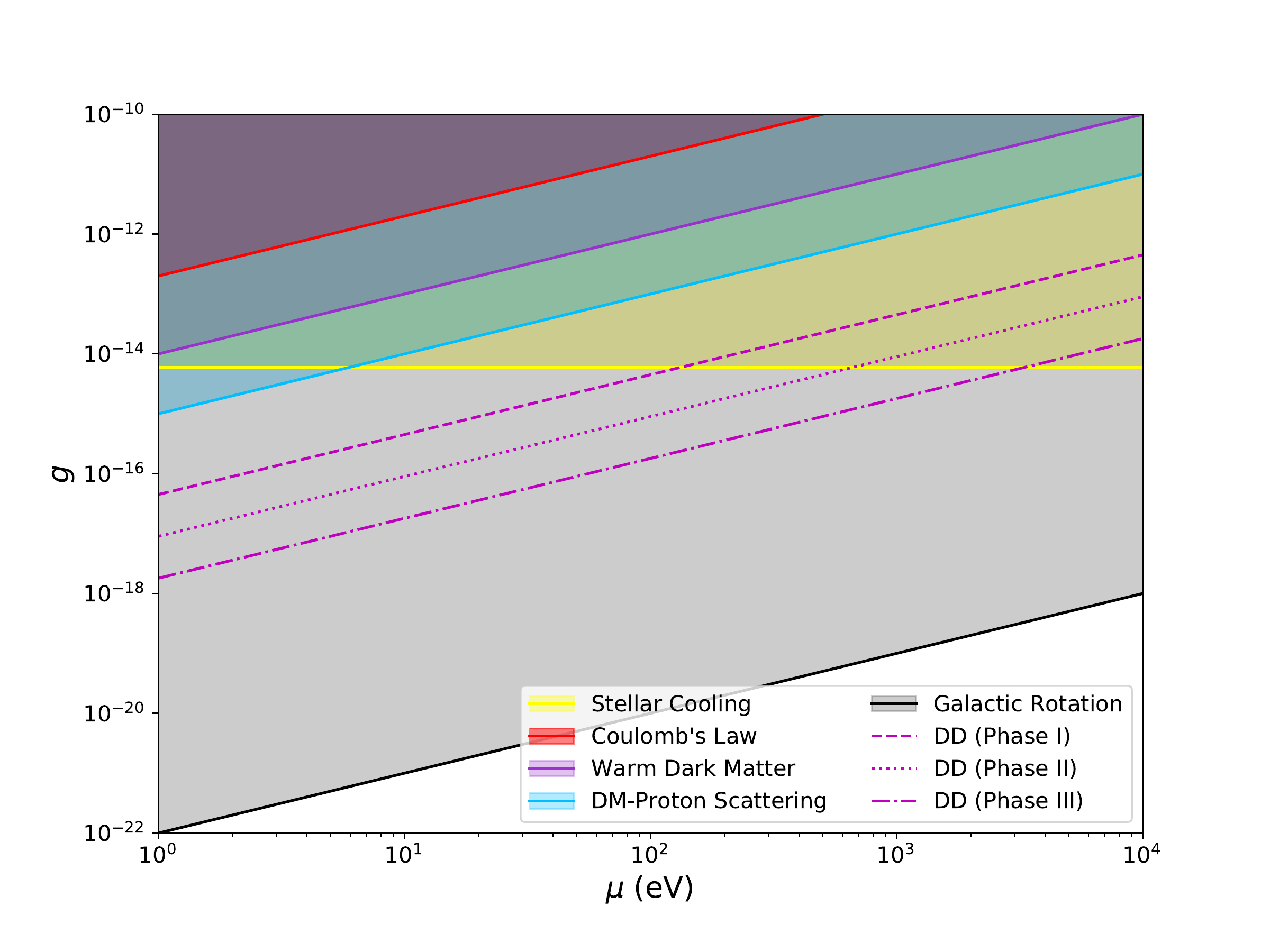}
\caption{Bounds on the direct coupling $g$ of the millicharged $\chi$ field to the photon as a function of $\mu$, as well as the projected reaches of a direct deflection experiment for the three sets of parameters discussed in \cite{PhysRevLett.124.011801}. Note that there is considerable uncertainty in the galactic spin-down bound; see \cite{Stebbins_2019}. Bounds and projected sensitivities for direct detection lie outside of the plotted region, at larger masses \cite{DD_KahnLin, DD_Essig}.}
\label{fig:MillichargeConstraints}
\end{figure}

A summary of the existing bounds on true millicharges, as well as projected sensitivities for a proposed direct deflection experiment (discussed below) are plotted in Fig. \ref{fig:MillichargeConstraints}.

\subsubsection{Constraints on Effective Millicharges Only}

\begin{figure}[htbp]
\centering
\includegraphics[trim=40 40 120 0, clip, width=1.0\linewidth]{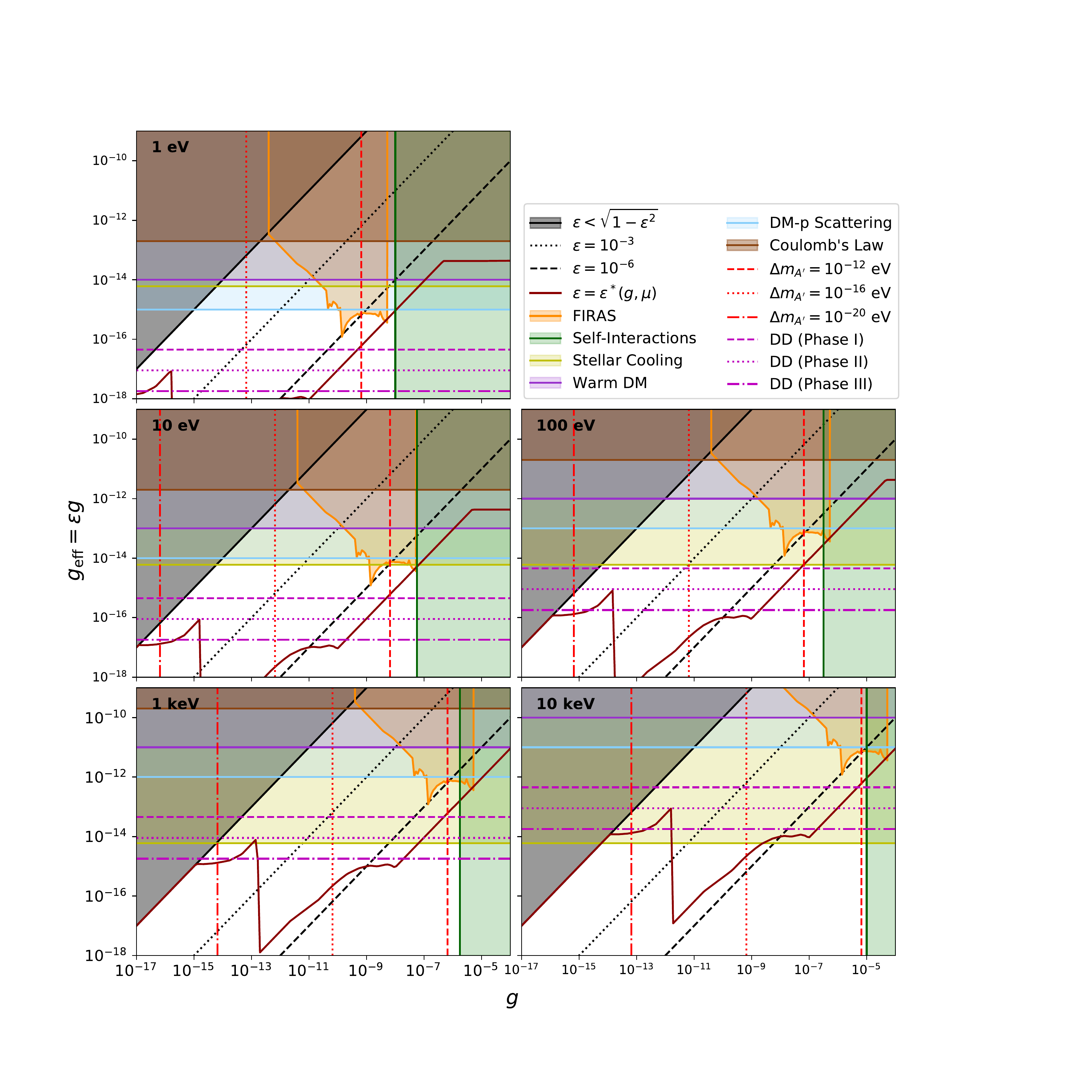}
\caption{Constraints on the possible values of $g_{\rm eff} = \epsilon g$ as a function of $g$ for a hidden photon-coupled $\chi$ field with masses from 1 eV to 10 keV. Shown are several existing bounds discussed in the text, as well as the projected reach of a direct deflection experiment \cite{PhysRevLett.124.011801}, including the three sets of parameters discussed in that reference. The FIRAS bound diverges at its high-$g$ edge, though this divergence is not shown as it is extremely narrow and is difficult to estimate numerically. Also shown are the values of $g$ corresponding to local hidden photon plasma masses of $10^{-12}$, $10^{-16}$ and $10^{-20}$ eV, which may correspond to additional constraints from black hole superradiance. Four lines corresponding to values of $\epsilon$ are included: $\epsilon/\sqrt{1-\epsilon^2}<1$ and $\epsilon=10^{-(3,6)}$ for reference (see e.g. \cite{essig2013dark} for a summary of current and expected future constraints on the hidden photon kinetic mixing), and $\epsilon=\epsilon^*(g,\mu)$ where $\epsilon^*(g,\mu)$ is the largest kinetic mixing allowed by those constraints at masses large enough to avoid plasma instabilities \cite{Lasenby_2020}. Note that none of these $\epsilon$ lines are current constraints, but they may lead to constraints if plasma instabilities are excluded in the future.}
\label{fig:DPConstraints}
\end{figure}

Several other bounds are specific to a hidden photon-coupled $\chi$ field. One such constraint comes from resonant conversion of photons to hidden photons, which would appear as a distortion in the CMB spectrum \cite{Mirizzi_2009}. Both photons and hidden photons have plasma masses that scale as $a^{-3/2}$ for most of cosmological history. However, during recombination, the decreasing fraction of unbound electrons causes the ratio of the Standard model photon and hidden photon plasma masses to fall. If this ratio crosses unity, CMB photons can potentially be adiabatically converted to hidden photons, distorting the CMB spectrum measured by, for example, the \textbf{Far Infrared Absolute Spectrophotometer (FIRAS)} \cite{Fixsen_1996}. Bounds on the coupling of $\chi$ can therefore be obtained from the agreement of FIRAS's measurements with a blackbody CMB spectrum.

This situation is complicated if hidden photons have a non-negligible ($\gtrsim 2\times10^{-14}$ eV) Proca mass alongside their plasma mass. In this case, the hidden photon's total effective mass will decrease only until the plasma mass falls below the Proca mass. This can also result in one or more energy crossings that distort the CMB spectrum.

The FIRAS bound shown in Fig. \ref{fig:DPConstraints} was made by substantially repeating the previous FIRAS hidden photon analysis \cite{Mirizzi_2009} using our redshift-dependent dark photon plasma mass \eqref{eq:DPplasmamass} in place of a constant dark photon mass. Note that the bound shown assumes that the hidden photon Proca mass is smaller than the current Standard Model photon plasma mass. For larger Proca masses, however, this bound changes because of the potential for new or different energy level crossings. In particular, the bound will extend to all values of $g$ less than or equal to those in the plotted bound, with a maximum value of $g_{\rm eff}/g$ that should be approximately that given in the existing FIRAS dark photon analysis \cite{Mirizzi_2009}.

Note that our analysis, like that of \cite{Fixsen_1996}, assumes a homogeneous plasma mass throughout the universe. The original hidden photon analysis of FIRAS's data has since been extended to account for plasma inhomogeneities \cite{PhysRevLett.125.221303}, but we do not attempt to reproduce this analysis for our model. We expect that accounting for inhomogeneities would somewhat broaden the parameter space excluded by FIRAS, but it should not significantly alter the conclusions of this work. Future measurements of the CMB spectrum may have the ability to probe substantial additional regions of millicharge parameter space, however \cite{kogut2019cmb}.

All of the preceding bounds on dark photon-coupled millicharges are plotted in Fig. \ref{fig:DPConstraints}, including both bounds on $g_{\rm eff}$---most of the bounds discussed above---and the self-interaction bound on $g$ itself. Also shown is an approximate bound coming from $\epsilon/\sqrt{1-\epsilon^2}<1$. The theoretical limit on $\epsilon$ is only $\epsilon<1$, which places no limit on $g_{\rm eff}/g$. However, order-unity values of $\epsilon$ correspond to large modifications to electromagnetism at scales larger than the inverse hidden photon mass, and are therefore constrained by various tests of Coulomb's law (see, e.g. \cite{PDGphoton}). Since the exact upper bound on $\epsilon$ will not affect our conclusions, we simply use $\epsilon/\sqrt{1-\epsilon^2}=1$ as an approximate bound for illustrative purposes.

There are additional bounds on hidden photon-coupled $\chi$ from \textbf{black hole superradiance}. In particular, hidden photons with (plasma) mass in the $10^{-13}-10^{-11}$ eV range could be superradiantly produced by stellar mass black holes, while those with masses below roughly $10^{-15}$ eV could be produced by supermassive black holes. As Fig. \ref{fig:DPConstraints} shows, this may rule out some additional sections of the considered parameter space, though we do not show precise exclusion bounds since they are still the matter of some debate in the literature (see, for example, \cite{Cardoso_2018, PhysRevD.96.035019}).

There may also be bounds on $\chi$ due to the consequences of \textbf{plasma instabilities} \cite{Lasenby_2020}. At sufficiently large couplings, dark matter plasma instabilities that appear when dark matter halos collide grow exponentially. Such exponential growth is generally not currently excluded by astrophysical observations, but it is likely to lead to signatures that are either detectable with further analysis of existing data, or will become detectable in the near future (at least if $\chi$ makes up all of dark matter; dark matter subcomponents are significantly more difficult to constrain \cite{PhysRevD.96.115021}). Plasma instabilities become significant for couplings $g$ (not $g_{\rm eff}$) above approximately
\begin{align}
    g \gtrsim 1\times10^{-22}\left(\frac{\mu}{1\text{ eV}}\right)\left(\frac{m_{A'}}{10^{-20}\text{ eV}}\right) \label{eq:plasmaInstabilities}
\end{align}
where $m_{A'}$ is the (Proca) mass of the hidden photon. This may include portions of otherwise unexcluded parameter space, depending on the hidden photon mass.

While these plasma instabilities are not currently excluded, future exclusion of them could lead to significant constraints on the parameter space we consider. In particular, \eqref{eq:plasmaInstabilities} would then set a lower bound on the hidden photon mass for a fixed value of $g$, which would in turn lead to a maximum value of $\epsilon$ due to existing bounds on hidden photons' kinetic mixing (see e.g. \cite{essig2013dark}). This is illustrated in Figure \ref{fig:MillichargeConstraints}, where the $\epsilon=\epsilon^*(g,\mu)$ line indicates the largest kinetic mixing allowed by existing bounds for any hidden photon heavy enough to avoid plasma instabilities, as set by \eqref{eq:plasmaInstabilities}. This line could therefore become an upper bound on $g_{\rm eff}$ if plasma instabilities are excluded in the future.

Several other known constraints on hidden photon models do not end up being relevant for the parameter space we are interested in. In particular, bounds from \textbf{solar luminosity measurements} and from \textbf{helioscopes} \cite{Redondo_2008} and from measurements of \textbf{galactic center gas clouds} \cite{PhysRevD.100.023001} are strictly weaker than those mentioned above.

\begin{figure}[t]
\centering
\includegraphics[width=0.6\linewidth]{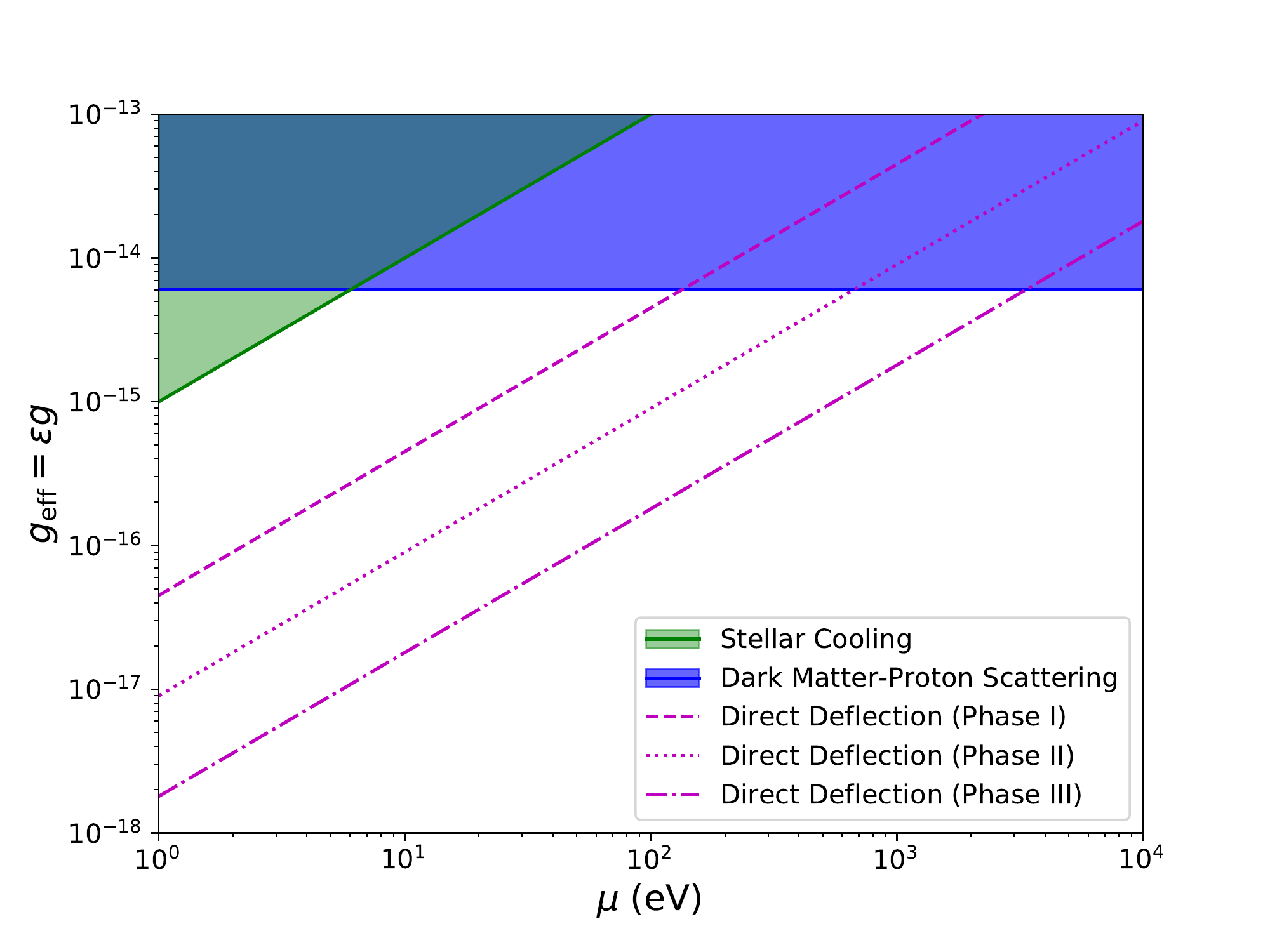}
\caption{The maximum value of $g_{\rm eff}$ subject to the constraints of Fig. \ref{fig:DPConstraints} as a function of $\mu$, as well as the projected reaches of a direct deflection experiment for the three sets of parameters discussed in \cite{PhysRevLett.124.011801}. Projected sensitivities for direct detection experiments lie outside of the plotted region, at larger masses and couplings \cite{DD_KahnLin, DD_Essig}.}
\label{fig:maxgeff}
\end{figure}

\subsection{Possibility of Detection}\label{sec:detection}

One exciting prospect for detecting the $\chi$ field if it is the main constituent of dark matter is direct deflection \cite{PhysRevLett.124.011801}. In a direct deflection experiment, the motion of millicharged dark matter relative to an oscillating magnetic field (with fixed position on Earth) results in a separation of the positively and negatively charged components in a region behind the magnetic field. This separation can then be detected using a resonant LC circuit inside of a superconducting shield: the shield prevents any other field sources from reaching the LC circuit, but the millicharged particles can flow through it unimpeded.

While no direct deflection experiments have been built as of writing, their projected sensitivity with only existing technologies already probes significant sections of millicharged dark matter parameter space with masses in the range $10^0-10^7$ eV and effective charges $g_{\rm eff}$ in the range $10^{-16}-10^{-9}$ (see Fig. 2 in \cite{PhysRevLett.124.011801}). Projections of longer-term work can potentially decrease the minimal charges by approximately a factor of 10. Note that these bounds weaken for hidden photon masses above approximately $10^{-7}$ eV, when the inverse hidden photon mass becomes comparable to the size of the experiment. 

While the upper part of this mass-charge parameter space is well motivated by freeze in models, the low-mass, low-coupling space has not previously been theoretically motivated. The $\chi$ field we have described, however, could exist in precisely this region. Direct deflection is therefore a promising approach to detecting or excluding it. We have included the detection limits for the three sets of experimental parameters described in \cite{PhysRevLett.124.011801} on Figs. \ref{fig:MillichargeConstraints}-\ref{fig:maxgeff}.

\section{Conclusion}

We have presented a new model of millicharged dark matter in which the cold relic abundance is the result of the arbitrary initial value of a field during the spontaneous breaking of a symmetry in the early universe, analogously to the well-known misalignment production of axion dark matter. This offers an alternative to the typical freeze-in and freeze-out dark matter models, which impose restrictions on possible millicharge masses and couplings in order to be consistent with cosmological observations.

Our model also motivates searching for millicharged dark matter at masses and couplings well below those preferred by previous freeze-in models. Notably, an experiment that is sensitive to this region of parameter space has already been proposed, allowing for potential detection or exclusion in parts of this space in the foreseeable future.

\appendix

\section{A Quark-Based Variant of Our Model}\label{app:quarkModel}

The Standard Model's matter content consists of fermions, rather than the scalar fields $\phi$. We might therefore wonder if we can construct an alternative version of our model based on fermionic fields.

One way to accomplish this is to consider a dark sector with a gauge group $U(1)\times SU(N)$ consisting of, at minimum, two Dirac spinors which we will refer to as up and down quarks since there will be no risk of confusion with the Standard Model analogues. We assume that this dark color group, like the Standard Model's color sector, has a non-zero vacuum expectation value of the quark bilinears
\begin{align}
    \left\langle \bar{u}u \right\rangle \sim \left\langle \bar{d}d \right\rangle \sim \Lambda_{\rm N}^3
\end{align}
with $\Lambda_{\rm N}$ the dark sector analogue of the QCD scale. If the quarks' masses are small compared to $\Lambda_{\rm N}$, then at high energies they admit the usual approximate $SU(2)_L\times SU(2)_R$ symmetry separately acting on the left- and right-handed components.

This leads to the usual spectrum of three low-energy pions at temperatures well below $\Lambda_{\rm N}$, when the $SU(2)_L\times SU(2)_R$ symmetry is broken to $SU(2)_V$: the $\pi^0$ and the $\pi^\pm$, where the latter have a $U(1)$ charge of magnitude equal to the difference of the quarks' charges. In the limit of zero $U(1)$ charge, these pions will have masses of
\begin{align}
    m_\pi^2 \sim \frac{\Lambda_{\rm N}^3}{F_{\pi}^2}(m_u+m_d)
\end{align}
just as in the Standard Model, with $F_\pi$ the dark sector analogue of the pion decay constant, presumably of order $\Lambda_{\rm N}$. The charged pions then acquire an additional mass due to radiative corrections, given by $g^2m^2$ times an $\mathcal{O}(1)$ factor, with $g$ the charged pions' $U(1)$ charge \cite{PhysRevLett.18.759, PhysRevD.55.7075}.

A particular case of interest is that of $m_u=m_d=0$. In this case, the $\pi^0$ is no longer produced, since ``misalignment'' of the $\pi^0$ field is actually just a change in the vacuum expectation values of the quark bilinears. However, the charges of the $\pi^+$ and $\pi^-$ cause them to still acquire a mass from radiative corrections.

The charged pions $\pi^\pm$ now reproduce the particle spectrum that we considered in the main text, with $v \sim \Lambda_{\rm N}$ and $\mu \sim m_\pi$. (If the quark masses are not zero, this also leads to an additional neutral pion.) In order for these pions to be sufficiently stable, we assume that there are no lighter particles (e.g. leptons) charged under the dark $U(1)$.

Both (or all three) pions will be produced via the misalignment mechanism when the temperature of the universe falls below $\Lambda_{\rm N}$, which should occur during inflation for the same topological reasons as for our scalar-only model. This does not occur for the Standard Model pions both because their mass is much larger than the value of Hubble at the QCD phase transition and because they rapidly decay.

In the case of massive quarks, one might worry that the decay of the neutral pion to two hidden photons could affect the cosmological history of the universe. To leading order, the decay rate of neutral pions to two hidden photons is \cite{NeutralPionDecay}
\begin{align}
    \Gamma(\pi^0\to\gamma'\gamma') = \frac{\alpha_D^2m_\pi^3N^2}{576\pi^3F_\pi^2}
\end{align}
with $\alpha_D=g^2/(4\pi)$ the dark fine-structure constant. Given our chosen range for $m_\pi$, the corresponding value of $v$, and our constraints on $g$, we have (for $N=3$)
\begin{align}
    \Gamma^{-1} \gtrsim 10^{37}\text{ s,}
\end{align}
which is sufficient to circumvent any bounds on dark matter stability \cite{PhysRevD.95.023010, Essig2013, Audren_2014, Poulin_2016}.

One could similarly worry about the conversion of charged pions to neutral pions through $\pi^+\pi^- \to \pi^0\pi^0$. This process is suppressed by $F_\pi^4$ \cite{Donoghue}, however, which is sufficient to make it insignificant.

\section{Effective Field Theory of \texorpdfstring{$\chi$}{Chi}} \label{app:EFTofChi}

In this appendix, we summarize the calculations of one loop radiative corrections to the $\phi$ Lagrangian discussed in Section \ref{sec:modelTheory}. Recall that, if the $SO(3)$ symmetry of the $\phi$ fields was exact, the most general possible Lagrangian for $\phi$ would be (up to dimension 4)
\begin{align}
    \mathcal{L}_{0} &= \frac{1}{2}(\partial_\mu\phi)^T(\partial^\mu\phi) + \frac{m^2}{2}\phi^T\phi - \frac{\lambda_\phi}{24}(\phi^T\phi)^2.
\end{align}
Now suppose that this symmetry is broken by the $U(1)$ gauge coupling at some high scale $\Lambda_{\rm UV} \gg m$. Then, at that scale, the complete Lagrangian for $\phi$ is
\begin{align}
    \mathcal{L}_{\rm UV} &= -\frac{1}{4}F_{\mu\nu}F^{\mu\nu} + \frac{1}{2}(D_\mu\phi)^T(D^\mu\phi) + \frac{m^2}{2}\phi^T\phi - \frac{\lambda_\phi}{24}(\phi^T\phi)^2.
\end{align}
At lower energies, however, the gauge coupling will lead to the appearance of three additional terms in the effective Lagrangian:
\begin{align}
    \delta\mathcal{L}_m &= - \frac{\delta\mu_\phi^2}{2}(\phi_x^2+\phi_y^2) - \frac{\delta\alpha}{24}\phi^T\phi(\phi_x^2+\phi_y^2) - \frac{\delta\beta}{24}(\phi_x^2+\phi_y^2)^2.
\end{align}
In general, there will also be running of the original Lagrangian parameters ($m$, $g$, $\lambda_\phi$), but we will not be concerned with these corrections.

\begin{figure}[t]
\centering
\includegraphics[width=0.6\linewidth]{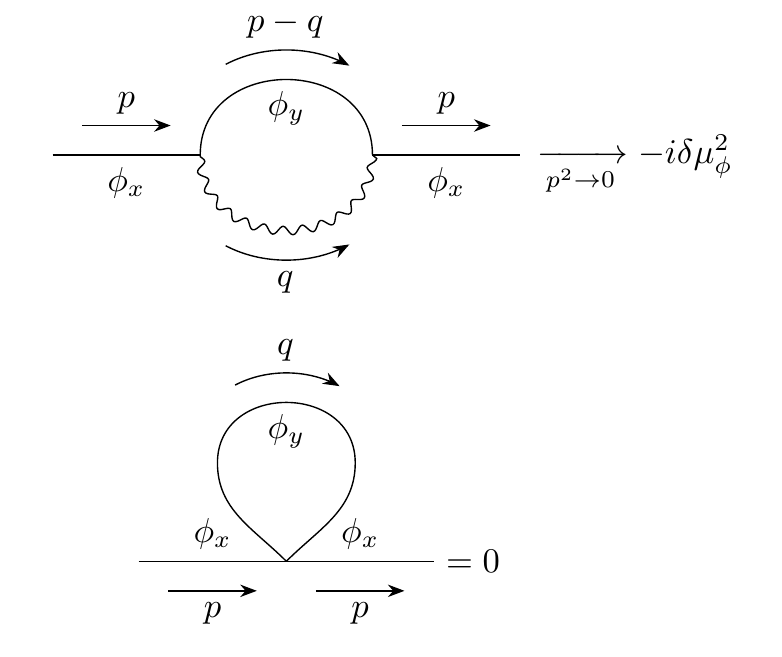}
\caption{The two one-loop diagrams contributing to the renormalization of the $\phi$ mass, $\mu_\phi^2$. The second diagram gives no logarithmic running; the first is evaluated in the text. Note that the relevant quantity is the diagram's zero-momentum limit, since we are interested in infrared effects; the momentum-dependent contributions lead to a field strength renormalization that does not contribute to the $\chi$ mass at lowest order.}
\label{fig:muDiagrams}
\end{figure}

There are two Feynman diagrams that could contribute to the mass renormalization $\delta\mu_\phi^2$, illustrated in Figure \ref{fig:muDiagrams}. The second diagram, however, lacks a mass scale, and therefore has zero logarithmic contribution to $\delta\mu_\phi^2$ (though its quadratic divergence corresponds to the EFT matching contribution to $\mu_\phi^2$). The remaining diagram (not including the external lines) is
\begin{align}
    -i\delta\mu_\phi^2 &= \left. -g^2\int\frac{\dd^4q}{(2\pi)^4}\frac{(2p-q)^2}{q^2((p-q)^2+m^2)} \right|_{p^2=0}.
\end{align}
Evaluating this using dimensional regularization with ultraviolet scale $\Lambda_{\rm UV}$ and keeping only the $p$-independent $\ln(\Lambda_{\rm UV}^2/m^2)$ terms, we have
\begin{align}
    \delta\mu_\phi^2 &= -\frac{g^2m^2}{16\pi^2}\ln\left(\frac{\Lambda_{\rm UV}^2}{m^2}\right).
\end{align}

\begin{figure}[t]
\centering
\includegraphics[width=0.8\linewidth]{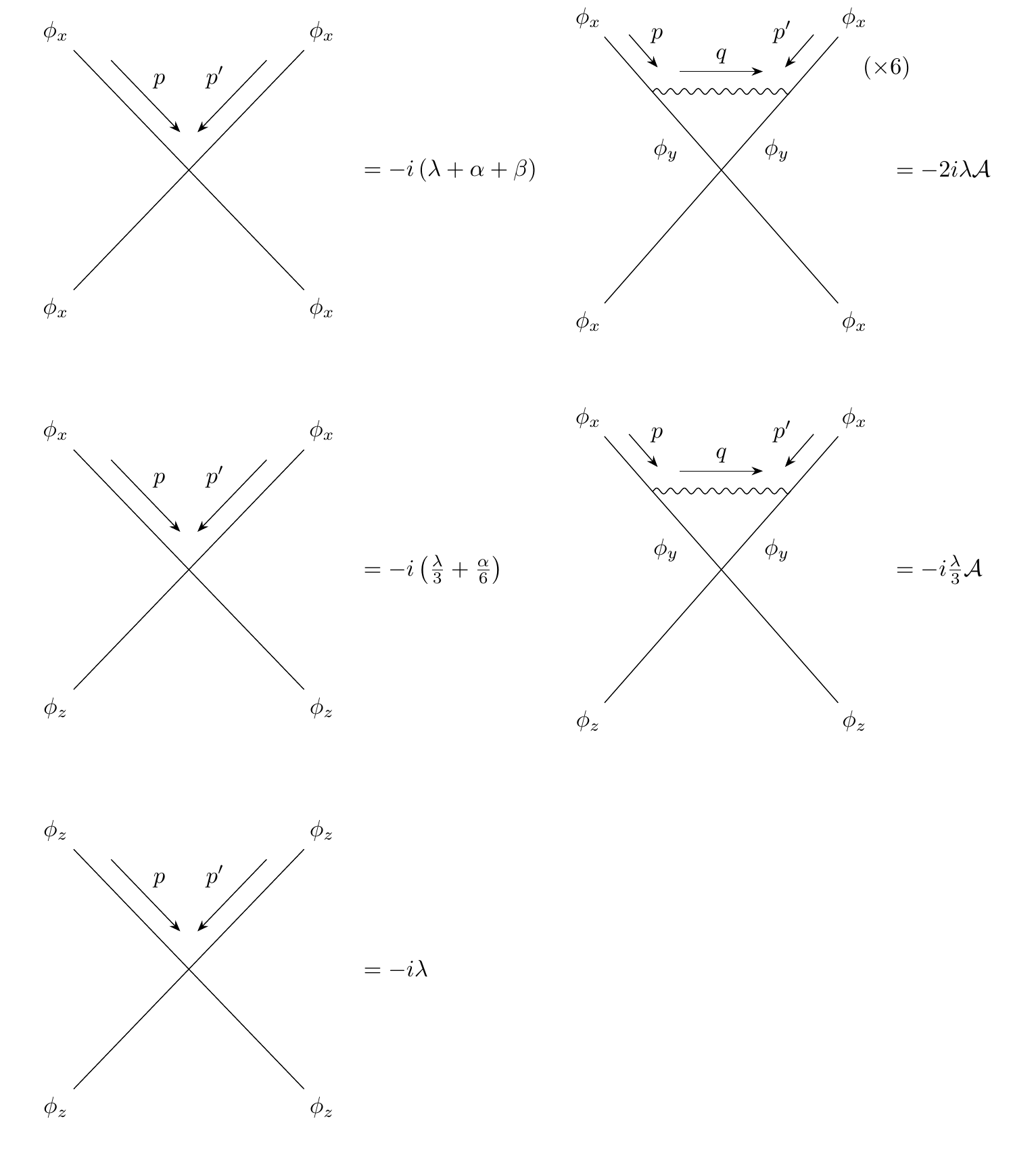}
\caption{A set of diagrams that can be used to evaluate the one-loop corrections to $\lambda$, $\alpha$ and $\beta$. The left column of diagrams shows the tree-level contributions of each term; note the combinatorial factors of $1/3$ that appear in vertices with mixed $\phi$ fields when using the normalization conventions in \eqref{eq:LphiFull}. The right column shows the one-loop diagrams that contribute to each vertex; note that there is no one-loop correction to $\phi_z^4$, since $\phi_z$ does not couple to the (hidden) photon. There are 6 copies of the diagram in the upper-right, with the photon loop between different pairs of external lines. In both one-loop digrams, $\mathcal{A}$ is an integral defined and evaluated in the text.}
\label{fig:alphaDiagrams}
\end{figure}

A number of diagrams contribute to the various $\phi^4$ vertices; these are tabulated in Figure \ref{fig:alphaDiagrams}, with the loop corrections written in terms of a common loop integral $\mathcal{A}$:
\begin{align}
    \mathcal{A} = ig^2\int\frac{\dd^4 q}{(2\pi)^4}\frac{(2p-q)\cdot(2p'+q)}{q^2[(p-q)^2+m^2][(p'+q)^2+m^2]}.
\end{align}
Again evaluating this using dimensional regularization with scale $\Lambda_{\rm UV}$ and keeping only the $p$- and $p'$-independent $\ln(\Lambda_{\rm UV}^2/m^2)$ terms, we have
\begin{align}
    \mathcal{A} = \frac{g^2}{16\pi^2}\ln\left(\frac{\Lambda_{\rm UV}^2}{m^2}\right).
\end{align}
Then, at one-loop level, we have
\begin{subequations}\begin{align}
    \delta\lambda &= 0 \\
    \delta\alpha &= \frac{g^2}{8\pi^2}\ln\left(\frac{\Lambda_{\rm UV}^2}{m^2}\right) \\
    \delta\beta &= 0.
\end{align}\end{subequations}
The zero correction to $\lambda$ is straightforwardly interpretable: $\lambda$ is the sole contribution to the $\phi_z^4$ vertex, which is unchanged by the $U(1)$ gauge coupling since $\phi_z$ is not charged. Understanding the zero correction to $\beta$ is slightly more involved: each $\phi^2A$ vertex corresponds to one insertion of $gR_z$ in an otherwise $\textrm{SO}(3)$-symmetric theory. Thus, one loop corrections can generate $\alpha$ terms of the form $\phi^T\phi(\phi_x^2+\phi_y^2) = \phi^T\phi(\phi^T R_z^2\phi)$, but they cannot generate a $\beta$ term $(\phi_x^2+\phi_y^2)^2 = (\phi^T R_z^2\phi)^2$, which requires four insertions of $gR_z$.

All of the loop calculations above were done in terms of the ultraviolet, unbroken $\phi$ fields. As discussed in the main text, the low energy spectrum of this theory is conveniently described in terms of a radial $\sigma$ field and the angular pions $\pi$ and $\pi^*$ (or $\chi$ and $\chi^*$). The Lagrangian for the light angular fields alone is given by \eqref{eq:chiExpanded}, but there are additional terms involving the heavy $\sigma$. For reference, we include all of the $\sigma$-dependent terms here, separated into the free $\sigma$ field terms and $\sigma$'s interactions with $\chi$:
\begin{align}
    \mathcal{L}_{\sigma,\rm{free}} &= \frac{1}{2}|\partial_\mu\sigma|^2 - m^2\sigma^2 - \frac{1}{6}\lambda_\phi v\sigma^3 - \frac{1}{24}\lambda_\phi \sigma^4 \\
    \mathcal{L}_{\sigma,\rm{int}} &= \frac{2}{v}\sigma|D\chi|^2 - \left(\frac{\alpha v}{3} + \frac{2\mu^2}{v}\right)\sigma|\chi|^2 + \frac{1}{v^2}\sigma^2|D\chi|^2 - \left(\frac{\alpha}{2}+\frac{\mu^2}{v^2}\right)\sigma^2|\chi|^2.
\end{align}

\acknowledgments

The authors would like to thank Robert Lasenby, Sebastian Baum, Peter Graham, Harikrishnan Ramani, and Michael Peskin for helpful discussions. ZB is supported by the National Science Foundation Graduate Research Fellowship under Grant No. DGE-1656518 and by the Robert and Marvel Kirby Fellowship and the Dr. HaiPing and Jianmei Jin Fellowship from the Stanford Graduate Fellowship Program. NT is supported by the U.S. Department of Energy under Contract No. DE-AC02-76SF00515.

\bibliographystyle{JHEP.bst}
\bibliography{main}

\end{document}